\documentclass[twocolumn,a4paper]{article}

\usepackage{latexsym}          
\usepackage[dvips]{graphicx}   
 \usepackage{amsmath}           
 \usepackage{amsfonts}          
 \usepackage{amssymb}           
\usepackage{pifont} 
\usepackage{bm}
\usepackage{hyperref}
\usepackage[dvips]{color}

\newcommand{\mysection}[1]{
\noindent {{\bf \rule{0pt}{14pt}#1}}}
\newcommand{\mycaption}[1]{{\small #1}}

\definecolor{myred}{rgb}{0.9,0,0}
\definecolor{mygre}{rgb}{0,0.5,0}
\definecolor{myblu}{rgb}{0.1,0.2,0.8}
\definecolor{mywhite}{rgb}{1,1,1}
\definecolor{myblack}{rgb}{0,0,0}
\definecolor{myoran}{rgb}{1,.4,.2}
\definecolor{mygra}{rgb}{.7,.7,.7}

\newcommand{\poly}[1]{(#1 \Vert)}

\newcommand{\bs}{\bm{s}}

\newcommand{\hex}{\textrm{hex}}
\newcommand{\sqoc}{\textrm{sq-oc}}

\newcommand{\Lam}{\Lambda}

\renewcommand{\and}{\wedge}

\newcommand{\triBu}{\blacktriangle}
\newcommand{\triBd}{\blacktriangledown}
\newcommand{\triWl}{\vartriangleleft}
\newcommand{\triWr}{\vartriangleright}

\newcommand{\be}{\begin{equation}}
\newcommand{\ee}{\end{equation}}

\renewcommand{\aa}{\alpha}
\newcommand{\bb}{\beta}
\newcommand{\cc}{\gamma}

\newcommand{\relaxx}{\mathcal{R}}
\newcommand{\projj}{\mathcal{P}}
\newcommand{\relaxxs}{\mathcal{R}_0}
\newcommand{\projjs}{\mathcal{P}_0}
\newcommand{\zv}{\vec{z}}

\newcommand{\wv}{\vec{w}}

\newcommand{\textcite}{\cite}

\begin{document}


\title{Deterministic Abelian Sandpile and square-triangle tilings}

\author{Sergio Caracciolo}
\author{Guglielmo Paoletti}
\author{Andrea Sportiello}

\date{\today}



\twocolumn[
\begin{center}
\Large{Deterministic Abelian Sandpile and square-triangle tilings}
\\
\rule{0pt}{18pt}%
\large{Sergio Caracciolo$^{\dagger}$,
Guglielmo Paoletti$^{\S}$
and
Andrea Sportiello$^{\dagger\,\ddagger}$,}
\\
\rule{0pt}{18pt}%
\small{$\dagger$
Dip.~Fisica, Universit\`a degli Studi di Milano, and INFN, via
G.~Celoria 16, 20133 Milano, Italy}
\\
\small{$\S$
LIP6, Universit\'e Pierre et Marie Curie,
4 Pl.\ Jussieu,
75252 Paris, France
}
\\
\small{$\ddagger$
LIPN, and CNRS,
Universit\'e Paris 13, Sorbonne Paris Cit\'e,
99 Av.\ J.-B.\ Cl\'ement, 93430 Villetaneuse, France}
\\
\rule{0pt}{18pt}%
{\tt sergio.caracci{}olo@mi.in{}fn.it}\;,
\quad
{\tt gugli{}elmo.paole{}tti@gma{}il.com}\;,
\quad
{\tt andre{}a.sp{}orti{}ello@lipn.fr}\;.
\\
\rule{0pt}{18pt}%
\today
\\
\vspace{5mm}
\begin{minipage}{.9\textwidth}
\small{{\bf Abstract:} 
The Abelian Sandpile Model, seen as a deterministic lattice automaton,
on two-dimensional periodic graphs generates complex regular patterns
displaying (fractal) self-similarity.  In particular, on a variety of
lattices and initial conditions, at all sizes, there appears what we
call an \emph{exact Sierpinski structure}: the volume is filled with
periodic patterns, glued together along straight lines, with the
topology of a triangular Sierpinski gasket.  Various lattices (square,
hexagonal, kagome,\ldots), initial conditions, and toppling rules show
Sierpinski structures which are apparently unrelated and involve
different mechanisms.  As will be shown elsewhere, all these
structures fall under one roof, and are in fact different projections
of a unique mechanism pertinent to a family of deterministic surfaces
in a 4-dimensional lattice. This short note gives a description of
this surface, and of the combinatorics associated to its construction.
}
\end{minipage}
\end{center}
\vspace{3mm}
]{}





\noindent
MSC 2000: Primary 52B20;
Secondary
37B15,
51M20.
\\
PACS 2010: 
05.65.+b\,, 
64.60.av\,,
89.75.Kd\,.


\mysection{\rule{0pt}{16pt}Introduction.}
Let $\Lam$ be the lattice in dimension 4, 
tensor product of two copies of the triangular lattice,
\hbox{$\Lam=
\langle e_1,e_2,e_3,e_4,e_5,e_6 \;|
\sum_{1 \leq i \leq 3}e_i = \sum_{4 \leq i \leq 6}e_i = 0 \rangle_{\mathbb{Z}}$}.
Consider the two-dimensional cell complex containing all the vertices
and edges of $\Lam$, and, as (oriented) faces, the triangles of the
two lattices and the parallelograms spanned by pairs $(e_1,e_4)$,
$(e_2,e_5)$ and $(e_3,e_6)$. We choose the orientation such that the
cycles $(e_1,e_2,e_3)$, $(-e_1,-e_2,-e_3)$ and
$(e_1,-e_4,-e_1,e_4)$ are upward faces,
and similarly with $(123)\to(456)$ and $(14)\to(25),(36)$.
We call a \emph{surface} a
connected and simply-connected collection of faces in the cell complex
above, with all upward faces.

Embeddings of $e_1, \ldots, e_6$ in $\mathbb{R}^2$ satisfying the
forementioned orientation constraints correspond to projections of the
4-dimensional cell complex on a 2-dimensional real space, such that
surfaces are mapped injectively.  An example of such an embedding is
$
(e_1, \ldots, e_6) = 
(\omega^3, \omega^{11}, \omega^7, \omega^0, \omega^8, \omega^4)
$,
where $\omega^k = (\cos \frac{k \pi}{6}, \sin \frac{k \pi}{6})$.
In this case, surfaces correspond to tilings of regions of the plane,
composed only of squares and triangles of unit sides, and along
directions multiple of $\pi/6$. These tilings are called
\emph{square-triangle tilings} in the literature.

Any other projection
is topologically equivalent, provided that $e_1+e_2+e_3 = e_4+e_5+e_6
=0$ and the orientation of the faces is preserved.  We call
\emph{valid} such a projection.  The set of valid projections is an
open portion of an algebraic projective variety.  We call
\emph{degenerate projections} those on the boundary of this open
set. Under degenerate projections, the image of some faces is a
segment or a point.

\newpage

A seminal work of de~Brujin for Penrose--Ammann lozenge tilings
\cite{debruj} has first illustrated the possibility that projections
of deterministic surfaces from a high-dimensional periodic
cell-complex could explain features of two-dimensional aperiodic
incommensurable tilings.  The square-triangle case discussed here
shows a similar phenomenon.

Square-triangle tilings have also distinguished properties, among
which is a relation with Algebraic Geometry, generalising the
well-known connection between lozenge tilings and Schur functions (see
e.g.\ \cite{Boro}).
The algebra of Schur functions has ubiquitous three-index structure
constants $c_{\lambda,\mu}^{\bar{\nu}}$, called
\emph{Littlewood--Richardson} (LR) coefficients \cite{LR}.
When the Young diagrams $\lambda$, $\mu$, $\bar{\nu}$ are boxed in a
rectangle $(d-n) \times n$ (as is the case, e.g., when they label
cells of the Schubert variety), there exists a relation
(\emph{Poincar\'e duality}) which acts as complementation at the level
of diagrams, $\nu \leftrightarrow \bar{\nu}$, and the LR coefficients
are symmetric in all three indices if the upper one is complemented,
$c_{\lambda,\mu}^{\bar{\nu}} =: c_{\lambda,\mu,\nu}$. As shown by
P.~Zinn-Justin \cite{pzj_LR}
and Purbhoo \cite{purb}, the LR coefficients correspond to the
enumerations of square-triangle tilings over triangoloids whose three
sides are built from $\lambda$, $\mu$ and $\nu$, respectively.  
Two degenerate projections of these surfaces
reduce to portions of the square and of the triangular
lattice. As degenerate projections transform some faces into segments
or points, the bijective correspondence is preserved only if extra
integer labelings, encoding the disappeared faces, are added to the
resulting structures.  These limiting tilings, together with the
auxiliary labelings, correspond to the original Littlewood--Richardson
rule \cite{LR} in the square case, and to the Knutson--Tao (discrete)
honeycombs \cite{KT1, KT2} in the triangular case.





\newpage

\mysection{ASM and square-triangle tilings.}
The purpose of this paper is to illustrate another unsuspected feature
specific of square-triangle tilings, namely of encoding the
\emph{exact Sierpinski structures} that arise in the Deterministic
Abelian Sandpile Model.  These structures have been identified on
various regular two-dimensional lattices, under various abelian
toppling rules, initial conditions and deterministic evolution
protocols, and square-triangle tilings describe them in a unified way.

The first occurrences of such structures have been presented, by the
authors, in \cite{guPhD,usEPL}, while the observation of approximated
versions of these structures (reproduced at a coarse-grained scale,
but locally deformed by some 1-dimensional defects) is much older
\cite{Osto}, and has first been made, only on the square lattice, for
the two most natural deterministic protocols: the evaluation of the
identity configuration in simple geometries 
\cite{DharManna, Kaplan, LeBRossin}, and the relaxation of a
large amount of sand put at the origin, in the (elsewhere empty)
infinite lattice \cite{DSC, HLMPW, amer, amer2}.

The `universal role' of the square-triangle tiling, in different ASM
realisations, should sound surprising, as the generic projection gives
incommensurable parallelogram-triangle tilings and does not live on a
discrete two-dimensional lattice, as is instead the case for the
sandpile models we consider. What comes out is that, in a remarkable
analogy with the mechanism discussed above for the combinatorics of
the Littlewood--Richardson rule and Knutson--Tao honeycombs, different
lattice ASM realisations occur at different ``rational'' points in the
set of valid projections (and its boundary, of degenerate
projections).

As this short paper is within a series, we do not give here an
introduction to the Abelian Sandpile Model. The interested reader can
consult the beautiful review by Deepak Dhar \cite{dharLN}, who first
established a large part of the theory.  For aspects of the model more
strictly related to the features discussed here, the reader can refer
to the PhD thesis of one of the authors \cite{guPhD}, or the shorter
papers \cite{usEPL} and \cite{usCipro}. Here we will only concentrate
on the aspects concerning the surfaces in the square-triangle tiling
corresponding to the exact Sierpinski structures in the ASM.

The sandpile configurations are height vectors $\zv=\{z_i\}$, with
variables $z_i \in \mathbb{N}$ associated to vertices $i$ of a graph
$G=(V,E)$.  There exists a notion of \emph{stable} configuration, and
a more restrictive notion of \emph{recurrent} one. \emph{Transient} is
a synonimous for non-recurrent. There exists a notion of
\emph{forbidden sub-configuration} (FSC), and a stable configuration
is recurrent iff it has no FSC. More generally, a configuration is
\emph{recurrent over $W \subseteq V(G)$} if it has no FSC contained
within $W$, thus making recurrency a local notion (like instability).
Local recurrency and instability are dual notions, if we set in the
wider frame of multitoppling ASM, as first shown in \cite{usCipro}.
The \emph{toppling matrix} $\Delta$ encodes the dynamics of the
sandpile, and determines a subdivision of $\mathbb{Z}^{V(G)}$ into
equivalence classes. There exists exactly one stable recurrent
configuration within each class.  Unstable configurations $\zv$ can be
\emph{relaxed} to stable ones, $\wv = \relaxx \zv$. Stable transient
configurations can be \emph{projected} to the unique recurrent
representative in the class, $\wv = \projj \zv$. The operators
$\relaxx$ and $\projj$ correspond to find the fixed point of
iterated maps, $\relaxxs$ and $\projjs$, corresponding to ``rounds''
of the procedure.\footnote{In $\relaxxs$ one can perform at most one
  toppling per site, in $\projjs$ one adds a single frame identity,
  and then relax.}

A number of structures and operations on square-triangle tilings can
be introduced, that will reproduce, under the various projection
procedures, the forementioned counterparts in the various ASM
realisations.  We dub all these features of the square-triangle
setting with the ``axiomatic'' attribute, as the reason for their
names emerges only when the projection procedure is explicitated.
Note that we are not able to reproduce \emph{all} the relevant
features of the sandpile model. In particular, we are not able to
reproduce the $a_i$ operators (nor their counterparts $a_i^{\dagger}$
defined in \cite{usCipro}).  The main things we are able to reproduce
are summarised by the following list
\begin{itemize}
\item The notion of (ASM-)equivalence of configurations is trivialised
  at the axiomatic level: two tilings are equivalent if they have the
  same boundary.
\item The axiomatic notion of FSC correspond to cycles in the tiling
  satisfying certain local rules.
\item We have an axiomatic notion of $\projjs$, consisting in a local
  deformation along the cycles of maximal FSC's (w.r.t.\ inclusion).
\item Similarly, we can certify that regions encircled by certain
  cycles will undergo a round of relaxation. This gives an axiomatic
  local notion of unstable subconfiguration
  (USC).\footnote{Corresponding to the \emph{waves of topplings}
    \cite{IKP94, IP98}.}
\item We have an axiomatic notion of $\relaxxs$, consisting in a local
  deformation along the cycles of maximal USC's (w.r.t.\ inclusion).
\item We have a recursive description of the Sierpinski structures at
  the axiomatic level. As these structures in the ASM
  determine the classification of patches and propagators in certain
  backgrounds \cite{guPhD}, this induces a corresponding
  classification of axiomatic patches and propagators.
\item A choice of vectors $e_1$,\ldots,$e_6 \in \mathbb{R}^2$, and of
  ``masses'' $\{m_{123}, m_{456}, m_{14}, m_{25}, m_{36}\}$ for the five types
  of tiles, induces a notion of density for the patches. This allows
  to state an axiomatic version of the Dhar--Sadhu--Chandra
  incidence formula, first introduced, for the ASM, in
  \cite{DSC}.
\end{itemize}






\mysection{Sierpinski structures.}
Let $\bs=(s_k,\ldots,s_1,s_0)$ be a finite string of positive
integers, and $n(\bs) = \sum_i 3^i s_i$.  A Sierpinski structure is
labeled by a string $\bs$, and $n(\bs)$ is its size.  Structures of
the same size are equivalent.

An abstract Sierpinski gasket of index $k$ is defined as follows. At
index 0, it is just a dark upward triangle. At index $k+1$, it is
obtained from the gasket at index $k$ by subdividing all dark upward
triangles into three dark upward and one light downward triangles, all
of half the side. Light triangles which are there at index $k$, will
remain unchanged at all $k'>k$.  A light triangle has index $k$ if it
first appeared in a gasket at index $k+1$.  A gasket of index $k$ has
$3^k$ dark triangles, and $3^h$ light triangles of index $h$, for
$0\leq h < k$.  See Fig.~1.

\begin{figure}[tb]
\begin{center}
\setlength{\unitlength}{28.5pt}
\begin{picture}(7.8,6.5)(-3.94,-2.35)
\put(-3.48,-3.05){\includegraphics[scale=1.216]{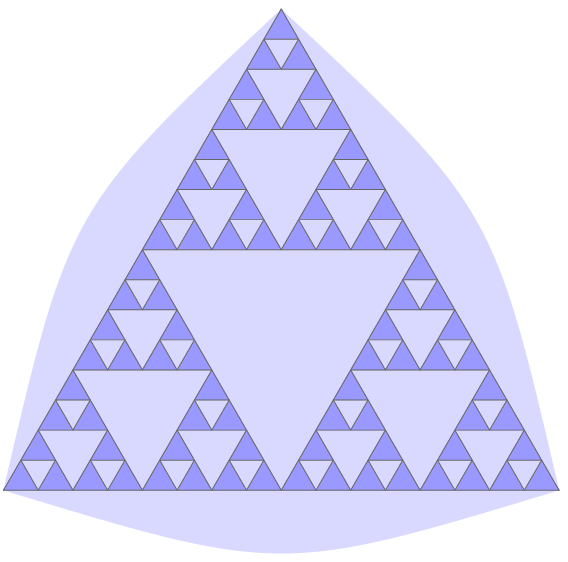}}
\put(-2.16487, 1.25){\makebox[0pt][c]{$a^{-1}$}}
\put(2.16487, 1.25){\makebox[0pt][c]{$b^{-1}$}}
\put(0, -2.5){\makebox[0pt][c]{$c^{-1}$}}
\put(0, 0){\makebox[0pt][c]{$1$}}
\put(1.73205, -1){\makebox[0pt][c]{$a$}}
\put(-1.73205, -1){\makebox[0pt][c]{$b$}}
\put(0, 2){\makebox[0pt][c]{$c$}}
\put(2.59808, -1.5){\makebox[0pt][c]{$aa$}}
\put(-0.866025, -1.5){\makebox[0pt][c]{$ba$}}
\put(0.866025, 1.5){\makebox[0pt][c]{$ca$}}
\put(0.866025, -1.5){\makebox[0pt][c]{$ab$}}
\put(-2.59808, -1.5){\makebox[0pt][c]{$bb$}}
\put(-0.866025, 1.5){\makebox[0pt][c]{$cb$}}
\put(1.73205, 0.){\makebox[0pt][c]{$ac$}}
\put(-1.73205, 0.){\makebox[0pt][c]{$bc$}}
\put(0, 3.){\makebox[0pt][c]{$cc$}}
\put(3.03109, -1.75){\makebox[0pt][c]{\small $aaa$}}
\put(-0.433013, -1.75){\makebox[0pt][c]{\small $baa$}}
\put(1.29904, 1.25){\makebox[0pt][c]{\small $caa$}}
\put(1.29904, -1.75){\makebox[0pt][c]{\small $aba$}}
\put(-2.16506, -1.75){\makebox[0pt][c]{\small $bba$}}
\put(-0.433013, 1.25){\makebox[0pt][c]{\small $cba$}}
\put(2.16506, -0.25){\makebox[0pt][c]{\small $aca$}}
\put(-1.29904, -0.25){\makebox[0pt][c]{\small $bca$}}
\put(0.433013, 2.75){\makebox[0pt][c]{\small $cca$}}
\put(2.16506, -1.75){\makebox[0pt][c]{\small $aab$}}
\put(-1.29904, -1.75){\makebox[0pt][c]{\small $bab$}}
\put(0.433013, 1.25){\makebox[0pt][c]{\small $cab$}}
\put(0.433013, -1.75){\makebox[0pt][c]{\small $abb$}}
\put(-3.03109, -1.75){\makebox[0pt][c]{\small $bbb$}}
\put(-1.29904, 1.25){\makebox[0pt][c]{\small $cbb$}}
\put(1.29904, -0.25){\makebox[0pt][c]{\small $acb$}}
\put(-2.16506, -0.25){\makebox[0pt][c]{\small $bcb$}}
\put(-0.433013, 2.75){\makebox[0pt][c]{\small $ccb$}}
\put(2.59808, -1.){\makebox[0pt][c]{\small $aac$}}
\put(-0.866025, -1.){\makebox[0pt][c]{\small $bac$}}
\put(0.866025, 2.){\makebox[0pt][c]{\small $cac$}}
\put(0.866025, -1.){\makebox[0pt][c]{\small $abc$}}
\put(-2.59808, -1.){\makebox[0pt][c]{\small $bbc$}}
\put(-0.866025, 2.){\makebox[0pt][c]{\small $cbc$}}
\put(1.73205, 0.5){\makebox[0pt][c]{\small $acc$}}
\put(-1.73205, 0.5){\makebox[0pt][c]{\small $bcc$}}
\put(0, 3.5){\makebox[0pt][c]{\small $ccc$}}
\end{picture}
\\
\setlength{\unitlength}{8.1pt}
\begin{picture}(24,12)(.1,-.4)
\put(-.2,-.2){\includegraphics[scale=.6]{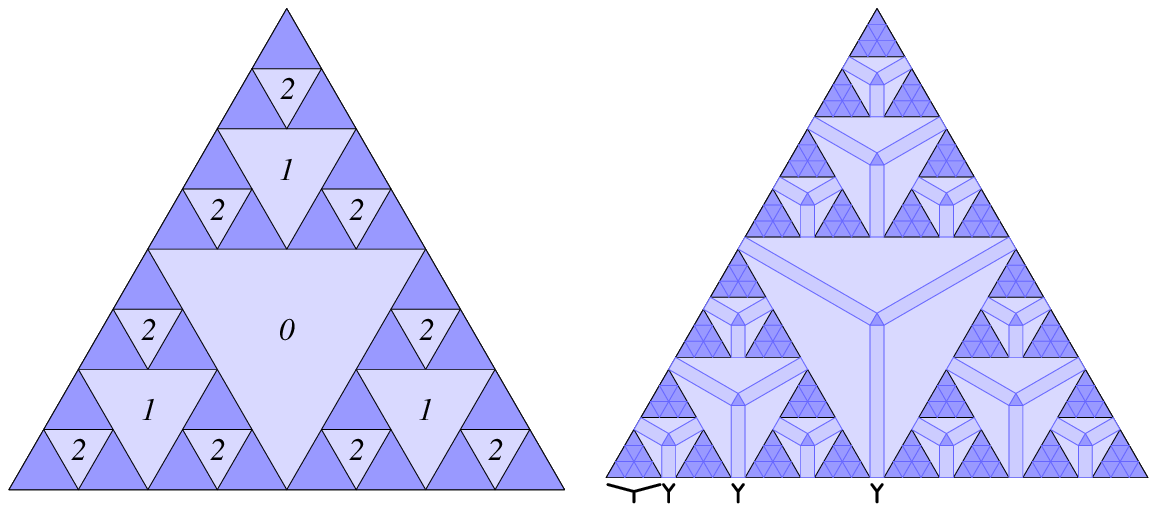}}
\put(12.5,-1){$s_3$}
\put(13.7,-1){$s_2$}
\put(15.2,-1){$s_1$}
\put(18.2,-1){$s_0$}
\end{picture}
\end{center}
\mycaption{%
{\small Fig.~1.
Bottom left: the Sierpinski gasket of index $k=3$. Bottom right: the structure of the patches, and the role of
  the parameter $\bs$. Top: the labeling of the light triangles.}}
\end{figure}

In the sandpile setting, the triangles of the gasket will determine
polygonal regions filled with a biperiodic patterns, called
\emph{patches} \cite{Osto}. Patches may be recurrent, transient or
marginal, depending on their behaviour under the burning test (see
\cite{usEPL}).


In a Sierpinski structure identified by $\bs$, all the dark triangles
correspond to transient patches, of triangular shape, with a side of
$s_k$ unit tiles.  Light triangles of index $h$ correspond to
polygonal regions filled with recurrent patches. These regions have
the aspect of triangoloids with concave sides, the sides being
polygonal lines composed of $2^{k-h}-1$ segments. The packing of unit
tiles depends in a certain fixed way on the integer $k-h$ and the
variables $s_{h'}$ for $h'>h$, and has no extra freedom, with an
exception: starting at the vertices of the triangoloids, we can have a
band of a patch with marginal density, of width $s_h-1$.\footnote{This
  corresponds to $s_h-1$ parallel \emph{type-I propagators},
  w.r.t.\ the definitions in \cite{usEPL, guPhD}.}  The three bands
meet at a triangular transient patch.

A transient patch contains a FSC only if ``sufficiently large'',
namely if it contains at least 7 unit tiles, packed in a shape
\hbox{\setlength{\unitlength}{10pt}
\begin{picture}(1.4,1)(.2,.3)
\put(0,0.4){$\circ$}
\put(0.23,0){$\circ$}
\put(0.23,.8){$\circ$}
\put(0.46,0.4){$\circ$}
\put(0.69,0){$\circ$}
\put(0.69,.8){$\circ$}
\put(0.92,0.4){$\circ$}
\end{picture}}.
Thus, a triangle of side up to 3 units filled with a transient patch, 
i.e.\ the shape
\hbox{\setlength{\unitlength}{10pt}
\begin{picture}(1.4,1)(.2,.3)
\put(0,0){$\circ$}
\put(0.23,0.4){$\circ$}
\put(0.46,0){$\circ$}
\put(0.46,0.8){$\circ$}
\put(0.69,0.4){$\circ$}
\put(0.92,0){$\circ$}
\end{picture}},
may still be part of an overall recurrent configuration. This has a
consequence on our Sierpinski structures: a structure with label $\bs$
is recurrent if and only if $1 \leq s_h \leq 3$ for all $h \leq
k$. These are the structures ultimately appearing in sandpile protocols.


Each region of the Sierpinski structure is filled with a periodic
pattern. The geometry of every region, including the number and
location of the unit tiles, is determined through a recursive
procedure.  Also the shape of the unit tiles, and their content in
terms of elementary squares and triangles, are determined
recursively. At this aim it is useful to introduce a labeling of the
regions of the Sierpinki gasket. We label the dark upward triangles
with words in the alphabet $\{a,b,c\}$, and the light downward
triangles with the same word as the dark triangle that originated
them. When a triangle of label $w$ is split, the three new triangles,
in the three directions, have labels $wa$, $wb$ and $wc$.  We also
give labels to the three external regions of the triangles, as
$a^{-1}$, $b^{-1}$ and $c^{-1}$.  See Fig.~1.


A triangle with label $w$ has three larger adjacent light triangles,
in the three directions, that have labels $\aa(w)$, $\bb(w)$ and
$\cc(w)$.  These three functions can be defined as follows.  Let
$\aa_w$, $\bb_w$ and $\cc_w$ the rightmost position along $w$ such
that, at its right, there are no more $a$, $b$ or $c$, respectively;
let us call $w|_{\ell}$ the truncation of $w$ to its first $\ell$
letters; let us understand that $a a^{-1} = b b^{-1} = c c^{-1} =
1$. Then $\aa(w) = w|_{\aa_w} a^{-1}$, and so on.


Complex tiles arise from the superposition of more elementary
ones. Only three tiles are indecomposable, and must be given as
input. These tiles correspond to the three square orientations in our
square-triangle tilings. The corresponding tilings appear outside the
triangle, at the three sides. The unit tile of label $w$ is composed
of the superposition of two copies of the tiles of labels $\aa(w)$,
$\bb(w)$ and $\cc(w)$. Unless $w=1$, one of these three words has
higher degree than the other (say $\aa(w)$). In this case, the six
tiles do not overlap, with the only exception that the two $\aa(w)$
tiles do overlap exactly on a $\aa(\aa(w))$ tile.  If $w=1$, no tiles
overlap. Each tile has 12 special positions along its boundary,
which determine the translation vectors of the recurrent, transient
and marginal tilings involving it, and the new tile inheritates its
owns positions from those of the three subtiles.

\newpage
\mysection{Dual tiles.}
Our construction in terms of the vectors $e_1$, \ldots, $e_6$ has a
number of covariances that allow to shorten our description
\\
\rule{0pt}{12pt}%
{\bf $C_3$-covariance} ($2\pi/3$ rotations):
\\
\rule{12pt}{0pt}
$(e_1,e_2,e_3,e_4,e_5,e_6) \to 
(e_2,e_3,e_1,e_5,e_6,e_4)$;
\\
{\bf exchange $\bm{(123)\leftrightarrow(456)}$} ($\pi/2$ rotations):\\
\rule{12pt}{0pt}
$(e_1,e_2,e_3,e_4,e_5,e_6) \to 
(e_4,e_5,e_6,-e_1,-e_2,-e_3)$;
\\
{\bf central symmetry} ($\pi$ rotations):\\
\rule{12pt}{0pt}
$(e_1,e_2,\cdots,e_6) \to 
(-e_1,-e_2,\cdots,-e_6)$.
\\
\rule{0pt}{12pt}%
We call \emph{polygon} a closed curve that is the boundary of some
square-triangle tiling. A polygon $P$ is determined by a cyclic
sequence in $\{1,\ldots,6,\b1,\ldots,\b6\}$, where $1$, $\b1$ stand
for $+e_1$, $-e_1$, and so on.  We use the shortcuts $\triBu$,
$\triBd$, $\triWr$ and $\triWl$ for the polygons $(123)$,
$(\b1\b2\b3)$, $(456)$ and $(\b4\b5\b6)$, respectively.

\begin{figure}[tb]
\begin{center}
\makebox[0pt][l]{\includegraphics[scale=1.2]{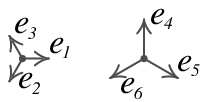}}
\includegraphics[scale=1.25]{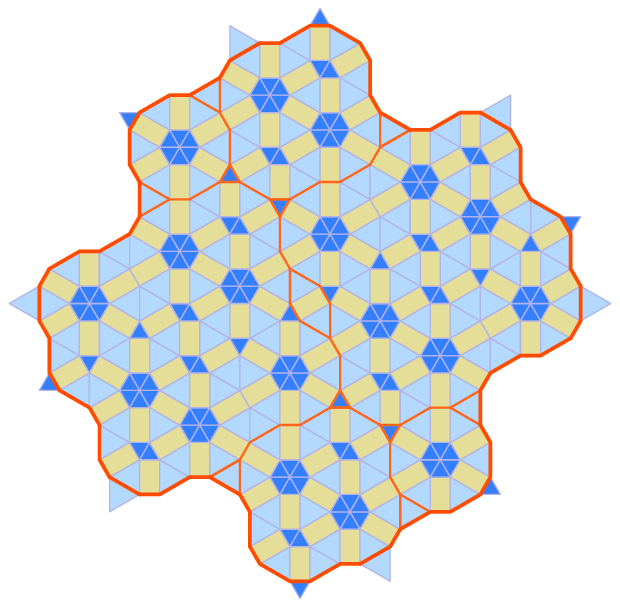}
\end{center}
\mycaption{%
{\small 
Fig.~2. The tile associated to $w=cba$. The interior
  orange lines describe the decomposition into $\aa(w)$, $\bb(w)$ and
  $\cc(w)$ tiles. The overlap, composed of a $\aa(\aa(w))$ tile and
  two light triangles, is in the middle.  The triangles outside
  the tile denote the 12 special positions.  Here we have
$u(P)=\big( (\b6 1),(\b2 \b6 1 \b6),(4 3 4 \b2 \b6 1 \b6 \b2),
(3 4), (\b5 3 4 3), (\b1 6 \b1 \b5 3 4 3 \b5) \big)$.
}}
\end{figure}

A centrally symmetric polygon $P$ is determined by a sequence of the
form $P=(i_1 i_2 \ldots i_k \underline{i_1} \,\underline{i_2} \ldots
\underline{i_k})$, where $\underline{\underline{i}}=i$. We use the
shortcut $\poly{i_1 i_2 \ldots i_k}$ in such a case.

A polygon $P$ is a \emph{dual tile} if both the
triple of polygons $(P,\triBu,\triBd)$ and the triple
$(P,\triWr,\triWl)$ (in these proportions) tile periodically the
plane. We call a \emph{transient\,/\,recurrent $\hex$ tiling} a tiling
of the two forms above, respectively.

The three fundamental parallelogram tiles are dual tiles.  The
dodecagon, $\poly{1\b6\b2 43\b5}$, is another example.  All the
tilings associated to dual tiles, except those deriving from the
fundamental parallelograms, have the topology of a \emph{hexagonal
  tiling}: each polygon $P$ is neighbour to other 6 $P$'s.  The
fundamental triangles are at the 6 triple points, with alternating
orientations cyclically along each $P$.



To each word $w$ as in the previous section can be associated a dual
tile $P(w)$, which is centrally symmetric.  The three fundamental
parallelograms are $\poly{41} = P(a^{-1})$ and so on. The dodecagon is
$\poly{1\b6\b2 43\b5}=P(1)$.

A pair of polygons $(P,Q)$ is a \emph{dual pair} if the sextuplet
$(P,Q,\triBu,\triBd,\triWr,\triWl)$ (in these proportions) tiles the
plane. We call a \emph{$\sqoc$ tiling} a tiling obtained as above.

Neglecting triangles (e.g., replacing them with $Y$-shapes), the
tiling has the square-octagon topology: any $Q$ tile is neighbour to 4
$P$ ones, and any $P$ tile is neighbour to 4 $P$'s and 4 $Q$'s,
alternating.\footnote{This fixes who's who among $P$ and $Q$.}  The
fundamental triangles are at the triple points of the square-octagon
topology. Each $P$ and $Q$ tile is adjacent to 8 and 4 triangles,
respectively, alternating dark\,/\,light, and, within dark and light
ones, of opposite orientations.

For each $w$, the pairs of tiles 
$\big( P(\aa(w)),P(w) \big)$, 
$\big( P(\bb(w)),P(w) \big)$ 
and $\big( P(\cc(w)),P(w) \big)$ 
are dual pairs.  For example, the dodecagon and any of the fundamental
parallelograms form a dual pair. 

Exceptionally, and analogously to what happens for $\hex$ tilings,
also all pairs of fundamental parallelograms are dual pairs, although
with a different topology, and with no ordering.

Each tile $P=P(w)$ appears in two hex tilings, three sq-oc tilings as
`octagon', and infinitely many sq-oc tilings as `square'. The union of
the positions of triple points among all these tilings has cardinality
12.  These 12 special positions break the perimeter of the tile into
open paths, related by the central symmetry. Thus, a list of 6
paths, $u(P)=(u_1,\ldots, u_6)$, determines simultaneously the
perimeter and the special positions, and
$P=\poly{1 u_1 \b6 u_2 \b2 u_3 4 u_4 3 u_5 \b5 u_6}$.

The recursive construction, at the level of these paths, leads to the
formulas (completed by $C_3$-covariance)
\begin{gather*}
(u_1 \b6 u_2)_w
=
(u_1 \b6 u_2 \b2 u_3)_{\bb(w)} \; \b6 \;
(\underline{u_6} 1 u_1 \b6 u_2)_{\cc(w)}
\\
\hspace*{-1mm}
\left\{ \rule{0pt}{20pt} \right.
\hspace*{-2mm}
\begin{array}{lc}
(u_1)_w = (u_1)_{\aa(w)} &
|\aa(w)|>|\bb(w)|
\\
(u_2)_w = (u_2)_{\bb(w)} &
|\aa(w)|<|\bb(w)|
\\
(u_1)_w = (u_2)_w = \varnothing &
\aa(w)=a^{-1}, \bb(w)=b^{-1}
\end{array}
\end{gather*}
The geometry of these paths is such that:\\
\;$\bullet$\ The sq-oct patches based on a
$(P,Q)$ dual pair may be adjacent to both recurrent and transient hex
patches, based both on $P$ and on $Q$, although with a restriction on
the direction of the (straight) boundary.\\
\;$\bullet$\ The hex transient tiling
based on $P(w)$ can be adjacent to the hex recurrent tiling based on
$P(w')$, if $w'$ is a prefix of $w$.

This ultimately leads to the consistency of the construction of the
Sierpinski structures (see Fig.~3).


\begin{figure*}[ht!]
\setlength{\unitlength}{20pt}
\begin{picture}(24,13.5)(-1,-.4)
\put(0,8){\includegraphics[scale=.48]{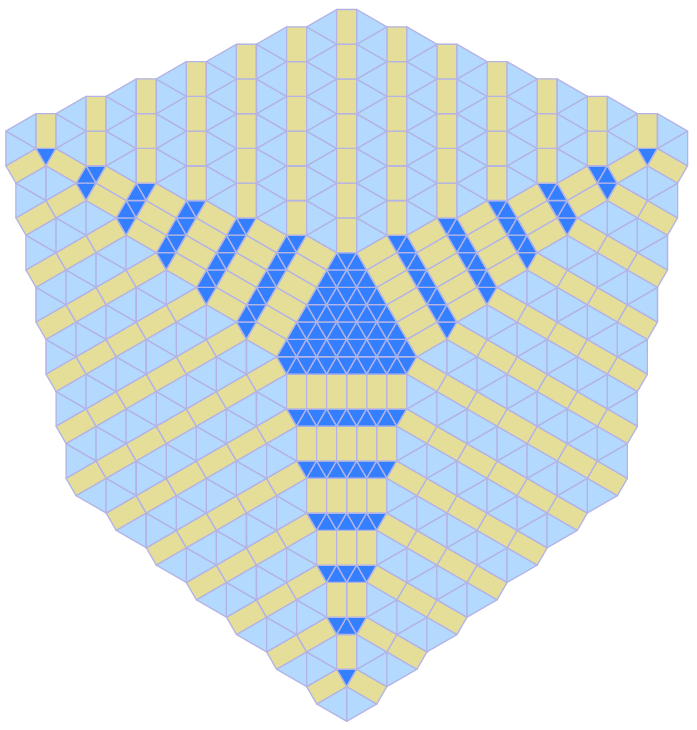}}
\put(4,4){\includegraphics[scale=.48]{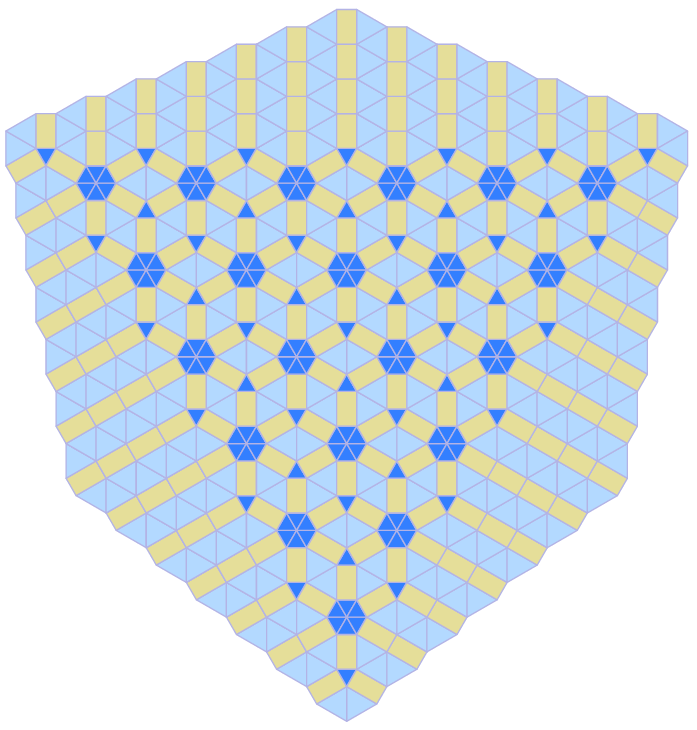}}
\put(0,0){\includegraphics[scale=.48]{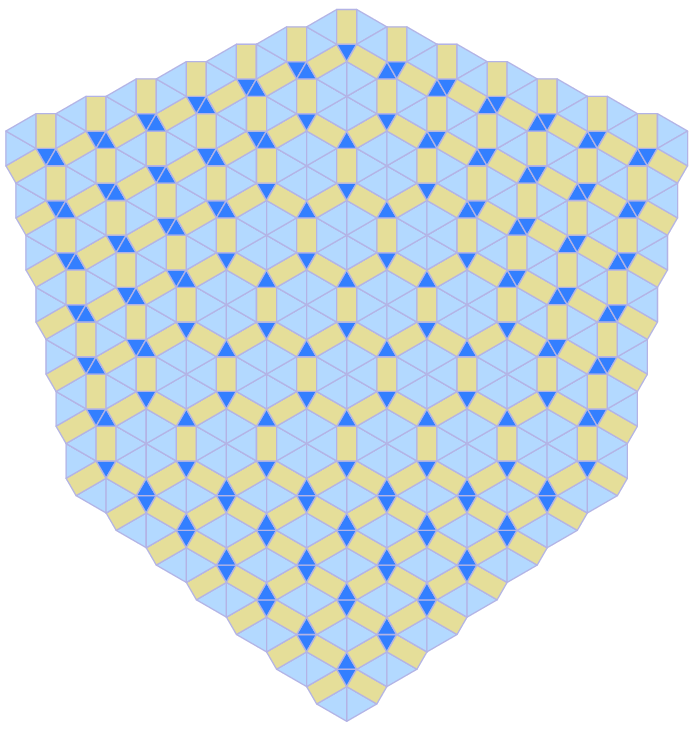}}
\put(9.5,0){\includegraphics[scale=.48]{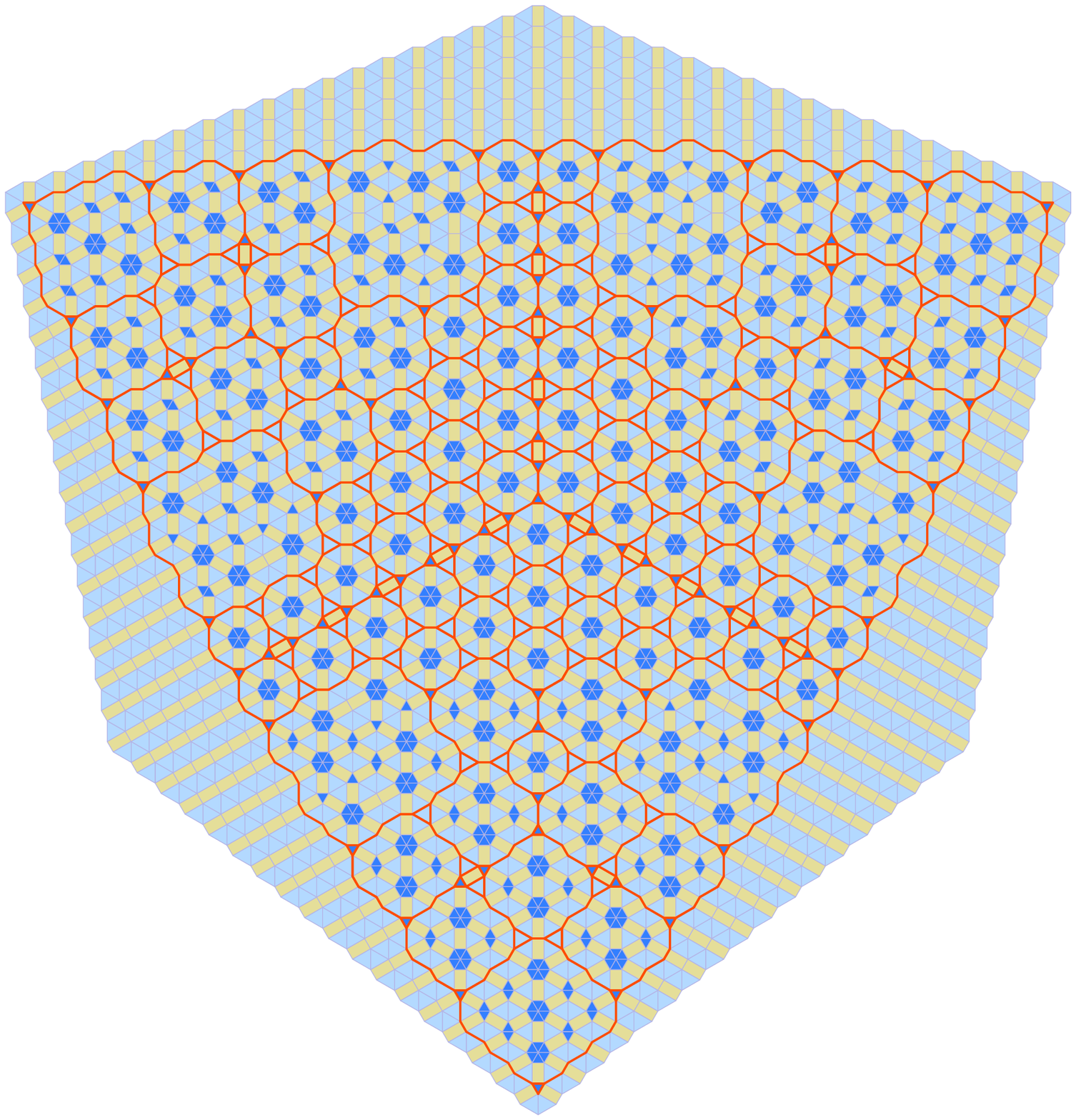}}
\end{picture}
\mycaption{%
{\small Fig.~3. Four classes of equivalent configurations: Left: three
  deterministic configurations, of size $n=6$. The two on top are
  stable but transient, and the one on the bottom is recurrent but
  unstable.  Applying $\projj$ and $\relaxx$, respectively, we obtain
  our axiomatic Sierpinski structure, (on the right at $\bs=(1,2,2)$,
  thus $n(\bs)=17$).  The patch structure is highlighted by the orange
  construction lines, showing the same topology of the Sierpinski
  gasket in Figure~1.}}
\end{figure*}

\newpage 

\begingroup
\renewcommand{\section}[2]{}%

\noindent
{\bf \large References.}

\end{document}